# Non-invasive imaging of solute redistribution below evaporating surfaces using $^{23}$Na-MRI

M.A. Chaudhry[1], S. Kiemle[2], A. Pohlmeier[1], R. Helmig[2], and J.A. Huisman[1]

[1] Agrosphere Institute (IBG 3), Forschungszentrum Jülich GmbH, 52428 Jülich, Germany.
[2] Institute for Modeling Hydraulic and Environmental Systems (IWS), University of Stuttgart, 70569 Stuttgart, Germany.

Corresponding author: Prof. Dr. Johan Alexander Huisman (s.huisman@fz-juelich.de)

**Key Points:**

- We used magnetic resonance imaging to experimentally capture downward driven density flow in porous media undergoing saline water evaporation
- Differences in salt distribution profiles and surface concentrations were observed between two media characterized by their different permeabilities
- Qualitative agreement between experimental observations and numerical simulations was seen for the evaporating systems.

**Abstract**

Saline water evaporation from porous media is a key phenomenon in the terrestrial environment and is linked to problems such as soil salinization and weathering of building materials. Recent modelling studies suggest the development of local instabilities due to density differences during evaporation in case of saturated porous media with high permeability. To experimentally investigate this and improve our understanding of near surface solute accumulation, we performed evaporation experiments on two types of porous media (F36 and W3) with intrinsic permeabilities that differed by two orders of magnitude. Using magnetic resonance imaging ($^{23}$Na-MRI), we monitored the development of solute accumulation and subsequent redistribution during evaporation under wicking conditions. The F36 sample showed an initial enrichment at the surface, but soon after a downwards moving plume developed that redistributed NaCl into the column. Average depth profiles of Na concentrations obtained from 3D imaging showed that the surface concentration reached only 2.5 mol $L^{-1}$, well below the solubility limit. In contrast, the W3 sample with lower permeability showed enrichment in a shallow near-surface zone reaching a concentration of over 6 mol $L^{-1}$. No fingering occurred although the mean evaporation rate was similar to that of F36 sand. Comparison of experimental results with numerical simulations using DuMu$^x$ for both samples showed qualitative agreement between measured and modelled solute concentrations. This study experimentally confirms the importance of density-driven redistribution of solutes in case of evaporating saturated porous media, carrying implications for predicting evaporation rates and the time to start of salt crust formation.

## 1. Introduction

Soil salinization is a process strongly entwined with evaporation (Shimojimaa et al., 1996) and causes significant soil degradation and losses in crop yield in semi-arid and arid regions with an estimated area of 932.2 million ha worldwide (Rodriguez et al., 1997; Chaves et al., 2009; Daliakopoulos et al., 2016). Salinization has both natural as well as anthropogenic origins and is determined by factors acting at various spatial-temporal scales (Schofield and Kirkby, 2003). Climatic conditions, geological formations, topography, oceanic influences, saline groundwater intrusions, and the presence of shallow water tables all play a role in naturally occurring salinization (Ghassemi et al., 1995).

Understanding evaporation of saline solutions from drying porous media is essential to understand soil salinization. Nachshon et al. (2011) presented a conceptual model for saline water evaporation where three stages were identified (SS1-SS3). In stage SS1, the evaporation rate is relatively high because the evaporation front is at the soil surface. Typically, a gradual decrease in evaporation rate is observed during the course of SS1. This is due to increased solute concentration at the sample surface, which results in a decreasing saturation vapor pressure as per Raoult's Law (Ho et al., 2006). In stage SS2, the concentration near the surface has increased to such an extent that precipitation is initiated, typically leading to a sharp decrease in evaporation rate. The process of crust formation on top (efflorescent) and within (subflorescent) the porous medium has been widely investigated in a range of studies (Eloukabi et al., 2013; Espinosa-Marzal and Sherer 2010; Nachshon et al., 2018; Weisbrod et al., 2013). In stage SS3, the evaporation rate is low because the evaporation front recedes into the porous media and/or due to pore blockage by dense salt crusts.

Soil salinization is further enhanced in case of a shallow water table because the continuous supply of water significantly sustains high evaporative water loss (Muwamba et al., 2018; Shokri and Salvucci, 2011). Previous studies have demonstrated that the depth to the water table is a key control on the evaporation rate, and that capillary connectivity of the water table to the surface is of high significance with respect to evaporation dynamics (Assouline and Or, 2014; Shokri and Salvucci, 2011). Less attention has been given to saline water evaporation in the presence of a shallow water table (Shokri-Kuehni et al., 2020), and the available studies often used experimental conditions where surface water content decreased with time (Shimojimaa et al., 1996; Rose et al., 2005; Jalili et al., 2011). This leads to a fundamentally different development of solute enrichment near the surface compared to a fully saturated evaporating soil.

To predict evaporation during stage SS1 and the onset of salt crust formation, it is important to accurately describe the development of near-surface concentration profiles due to evaporation. Elrick et al. (1994) derived analytical solutions of the convection-dispersion equation describing solute enrichment near the surface assuming a constant but depth-dependent water content profile. Their analysis predicts an enrichment of solute at the surface for homogenous porous media with exponentially decreasing concentrations with depth. More recent modelling work by Bringedal et al. (2022) assumed a fully saturated, homogeneous and isotropic porous medium undergoing evaporation in the presence of NaCl solute. They investigated the conditions in which a denser and gravitationally unstable saline solution may form near the evaporative surface due to enrichment (Geng and Boufadel, 2015; Wooding et al., 1997a). They showed that the intrinsic permeability crucially impacts the generation of such instabilities, and found that samples with high permeability ($10^{-10}$ $m^2$ to $10^{-13}$ $m^2$) are more prone to instabilities than samples with lower permeability. The downward redistribution of solutes due to density-driven flow would have implications for predictions of the evaporation rate and the initiation of crust formation (Shokri-Kuehni et al., 2017; Piotrowski et al., 2020; Li et al., 2022), the formation of saline groundwater below playas (Wooding et al., 1997b; Hamann et al., 2015), as well as salt polygon formation (Lasser et al., 2023). So far, this downward redistribution of solutes against the evaporative flux has been theoretically predicted and qualitatively visualized in Hele-Shaw cells (Wooding et al., 1997a). It has not yet been experimentally observed on a three dimensional scale.

Three-dimensional experimental investigations of flow instabilities require non-invasive imaging techniques, such as magnetic resonance imaging (MRI) or X-ray computed tomography (XRCT). MRI has many applications in material and geosciences (Blümich, 2000; Blümich et al., 2014; Haber-Pohlmeier et al., 2018; Haber-Pohlmeier et al., 2019). In the context of imaging flow instabilities, Dickson et al. (1997) applied $^1H$ MRI imaging to visualize viscous fingering in chromatography using a doping agent. Rose and Britton (2013) also used $^1H$ MRI to visualize viscous fingering in a reactive system with a rheological change at the interface between two fluids.

However, $^{1}H$ MRI is not suitable for investigating density-driven redistribution of Na in a single fluid without a doping agent that may interfere with the investigated flow processes. XRCT has also been used to investigate miscible and immiscible fluid displacement of oil in porous media by Berg et al. (2010) and Riaz et al. (2007), respectively. Suekane et al. (2017) used conventional XRCT to image how NaCl brine and Glycerol displaced water through viscous fingering in a saturated porous media doped with $CaI_2$. Although conventional XRCT offers high resolution imaging, the approach is not sufficiently sensitive for imaging minor density differences resulting from variable salt concentrations. Synchrotron XRCT on the other hand offers superior imaging quality and fast measurements as was demonstrated by Shokri-Kuehni et al. (2018) for their evaporating porous media with $CaI_2$ solution. However, the reliance on iodide salts makes it difficult to transfer the results to natural systems.

An interesting option for imaging solute redistribution during evaporation is $^{23}Na$ MRI (Werth et al., 2010). This approach has so far been mainly used to investigate plants and root-soil interactions. For example, Olt et al. (2000) obtained high resolution images (156 µm × 156 µm) of the Na distribution in plants. Quantification was complicated due to suboptimal relaxation times of Na inside different tissue bodies in relation to the minimum echo time of 3.2 ms for Na in their MRI system (Olt et al., 2000). Rokitta et al. (2004) obtained quantitative imaging results for Na concentration in stem tissue of salinity-resistant plants using in-situ reference tubes with a known Na concentration. More recently, Perelman et al. (2020) used $^{23}Na$ MRI to monitor NaCl accumulation near the soil-root interface of tomato plants. As in Rokitta et al. (2004), calibration using a reference tube was used to obtain quantitative information on Na concentration down to 0.01 mol $L^{-1}$. These results highlight the capability of $^{23}Na$ MRI for detecting small concentration differences in Na.

Within this context, the aim of this study is to use $^{23}Na$ MRI to investigate density driven flow during evaporation in two porous media with contrasting intrinsic permeability supplied with water from below (wicking). To achieve this aim, existing $^{23}Na$ MRI protocols were adapted to monitor Na redistribution with sufficiently short acquisition times. The $^{23}Na$ MRI measurements were accompanied with $^{1}H$ imaging to verify that the samples remained fully saturated during the experiments. The intrinsic permeabilities of the two samples were selected to obtain distinct differences in near-surface solute accumulation based on the work of Bringedal et al. (2022). Finally, experimental results were compared with numerical simulation obtained with the DuMu$^x$ model, which is able to simulate density-driven redistribution of solutes.

## 2. Basic Principles of MRI

MRI is based on the magnetic properties of atomic nuclei. In brief, many atomic nuclei possess a property termed spin linked to a microscopic magnetic moment. When placed in an external magnetic field, such spin systems adopt different energy states. Transitions to higher energy states are possible by resonant absorption of electromagnetic radiation, followed by subsequent emission of a signal with a strength $S_0$ and a temporal relaxation characterized by $T_1$ and $T_2$ relaxation times. Target nuclei such as $^1H$ and $^{23}Na$ are specifically addressed by the choice of the resonance frequency $v_0$, e.g. 200 MHz for $^1H$, and 52 MHz for $^{23}Na$ in a main magnetic field of 4.7 T. The resulting signal can be manipulated by additional radiofrequency pulses and magnetic field gradient pulses to make it sensitive to $S_0$ or probe the relaxation decay by variation of experimental parameters, mostly echo time, $t_E$ or repetition time $t_R$. Furthermore, magnetic field gradient pulses allow to encode the signal spatially. Many different pulse sequences exist. For practical applications in natural porous media, spin-echo sequences were found to be most convenient. The signal intensity in a voxel $S(r,t)$ as function of time is given by:

$$S(r,t) = S_0(r)\, exp\left(-\frac{t_E}{T_2}\right)\left[1 - exp\left(-\frac{t_R}{T_1}\right)\right], \tag{1}$$

where $T_1$ and $T_2$ are the medium-specific transverse and longitudinal relaxation times, and the echo time $t_E$ and the repetition time $t_R$ are experimental parameters adjustable by the operator. For different media, these parameters may need to be adjusted. Ideally, the echo time ($t_E$) is significantly shorter than the $T_2$ decay time and the repetition time ($t_R$) is five times longer than the $T_1$ decay time. In this case, the signal intensity of a voxel is directly proportional to the volume concentration of the investigated nuclei in the voxel. For more information about MRI, the reader is referred to standard textbooks (e.g. McRobbie et al., 2017; Blümich, 2000).

## 3. Materials and Methods

### 3.1 Evaporation Experiments

F36 and W3 sands (Quarzwerke Frechen, Frechen, Germany) were selected as porous media. Table 1 summarizes key properties of the selected materials. F36 has a more uniform particle size distribution compared to W3. Hydraulic parameters for the two materials were obtained by preparing three samples each with a wet packing approach using metal cylinders with a diameter of 5.3 cm and a length of 5.1 cm. These six samples were placed on a sand bed and water loss was recorded after subjection to a range of negative pressures between 5 mbar and 300 mbar. After this, the samples were oven-dried at 60 °C and the final dry mass was recorded. The Brooks-Corey equation was fitted to the obtained water retention data to obtain the hydraulic parameters $\theta_{res}$, $\theta_{sat}$, $p_b$, and $\lambda_{BC}$ (Table 1). A second set of samples of F36 and W3 were used to determine saturated hydraulic conductivity using the falling head method, after which the intrinsic permeability $k$ was determined using (Kasenow, 2002):

$$k = \frac{K_s\,\mu}{g\rho} \quad (2)$$

where $K_s$ is the saturated hydraulic conductivity, $\mu$ is the fluid viscosity (1.00 mPa s at 20°C), $\rho$ is the fluid density (0.998 g $cm^{-3}$ at 20°C), and $g$ is the gravitational acceleration (9.8 m $s^{-2}$). This resulted in a mean intrinsic permeability of $7.83 \times 10^{-12}$ $m^2$ for F36 and $3.2 \times 10^{-14}$ $m^2$ for W3. Based on the work of Bringedal et al. (2022), it is anticipated that F36 will show gravitationally instable solute enrichment and density-driven downward flow, whereas this is not expected for W3.

For MRI imaging, custom-made metal-free sample holders were used (Figure 1). The sample holders were made from polymethylmethacrylate (PMMA) with an inner diameter of $d$ = 3.1 cm, a total length of $l$ = 14 cm, a maximum filling height of $h_{max}$ = 10 cm and a bottom filter plate with a thickness of 0.3 cm (P3, Robu, Hattert, Germany). To prepare the packings, the sample holder was filled with 1.0 M NaCl solution to a height of 2 cm. Subsequently, sample material was added to an intended filling height ($h$) of 4 cm and compacted by gentle tapping. Excess saline solution was allowed to drain out through the bottom outlet, after which the total mass of the sample material and water in the sample was determined from the mass balance of added and removed amounts of sample material and water. The initial volumetric water content, $\theta$, was then calculated using:

$$\theta = \frac{m_{sol}}{\rho_{sol} V_{packing}} \quad (3)$$

where $m_{sol}$ is the mass of solution in the sample holder, $\rho_{sol}$ is the density of the 1.0 mol $L^{-1}$ NaCl solution (1.039 g $cm^{-3}$; Simion et al., 2015), and $V_{packing}$ is the bulk volume of the saturated sand packing.

Before inserting the samples into the MRI scanner, the bottom inlet was connected to an external reservoir with 1 mol $L^{-1}$ NaCl solution with a plastic tube that was initially closed with a valve. This reservoir was placed on a balance (accuracy ±0.01 g, EX2202, Ohaus Corporation, Parsippany, USA) positioned on a table with adjustable height outside of the MRI scanner (Figure 1a). A concentration of 1 mol $L^{-1}$ was used here to provide a suitable range between the initial concentration and the solubility limit of NaCl, which is 6.14 mol $L^{-1}$ at 25 °C (Sanz and Vega, 2007).

To start MRI imaging, the sample was placed in the $^{23}Na/^{1}H$ probe head with 66 mm internal diameter (Bruker Mini SWB 90 imaging probe DR). This probe head was subsequently inserted in the MRI scanner (Bruker super wide bore scanner at $B_0$ = 4.7 T, Figure 1a). The water table in the reservoir outside of the MRI was adjusted to a height of 2 cm below the sample surface. As low suction led to negligible water loss for both sands during the determination of the water retention curve, full saturation of the sand is expected at the start of the experiment. After positioning the sample in the MRI, the valve between the sample and the reservoir was opened and the sample was allowed to equilibrate for 1 hour. After this, a dry-gas diffuser was placed immediately above the column inside the MRI scanner to generate a turbulent stream of $N_2$ gas (Figure 1b) and evaporation was started by switching-on the gas flow with a rate of 60 $dm^3$ $h^{-1}$. Ambient conditions within the laboratory were monitored with a multi-purpose probe (FHAD 46-C2, ALMEMO® D06, Ahlborn, Holzkirchen, Germany). The mean temperature was 20°C ± 2°C and the mean relative humidity was 40% ± 5%. During evaporation, the mass of the reservoir was recorded hourly to determine the evaporation rate $E_{rate}$ of the sample using:

$$E_{rate} = \frac{\Delta m}{\rho_{sol} A_{evap}.t_{diff}} \quad (4)$$

where $\Delta m = m_1 - m_2$ is the mass difference during the time interval $t_{diff} = t_1 - t_2$, $\rho_{sol}$ is the solution density, and $A_{evap}$ is the evaporative area.

Table 1. Properties of the two materials used for the evaporation experiments. Particle size ranges were provided by the supplier (Quarzwerke Frechen). The parameters of the Brooks-Corey model were obtained by fitting to measured water retention data.

| Sand | F36 | W3 |
|---|---|---|
| average particle size (µm) | 160 | 90 |
| particle size range (µm)<br>type (USDA) | 90 – 355<br>medium sand | 2 – 400<br>very fine sand |
| intrinsic permeability ($m^2$) | $7.83 \times 10^{-12}$ | $3.2 \times 10^{-14}$ |
| Brooks-Corey saturated water content ($\theta_{sat}$) | 0.39 | 0.31 |
| Brooks-Corey residual water content ($\theta_{res}$) | 0.02 | 0.08 |
| entry pressure ($p_b$ in Pa) | $4.69 \cdot 10^3$ | $1.12 \cdot 10^4$ |
| Brooks-Corey parameter $\lambda_{BC}$ | 5.04 | 0.384 |

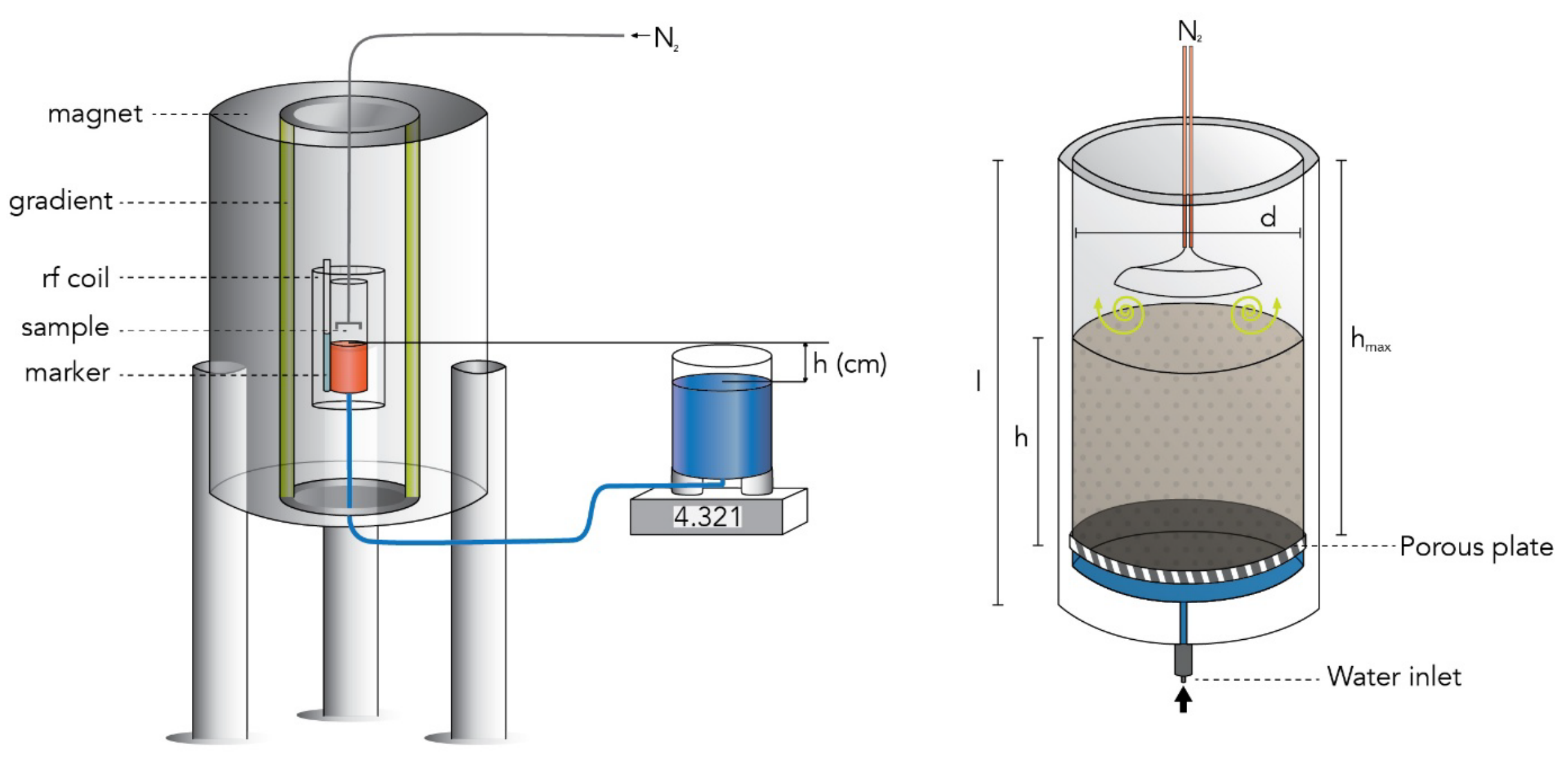

(a) (b)

Figure 1. Illustration of the experimental set-up with a) MRI device and the overall arrangement of the sample and reservoir, b) sketch of the sample holder with total length l, maximum fillable height $h_{max}$, the filled height h, and diameter d. A supply of water is provided to the bottom inlet marked with an arrow and the nitrogen gas stream is injected from the top inlet and circulated over the sample.

### 3.2 Magnetic Resonance Imaging

All MRI experiments were performed using a Bruker super wide bore scanner with $B_0$ = 4.7 T. This results in Larmor frequencies of $\nu_0$ = 200 MHz for $^1$H and 52.9 MHz for $^{23}$Na. The MRI scanner is equipped with a 3D orthogonal gradient/shimming system delivering a maximum gradient strength of 700 mT m$^{-1}$. The scanner was operated by an Avance III console and controlled by Paravision software (version 6, Bruker Microimaging, Rheinstetten, Germany). Table 2 summarizes the different MRI protocols used for mapping Na content using $^{23}$Na imaging and water content mapping using $^1$H imaging. Different $^{23}$Na-MRI imaging protocols were used for F36 and W3 samples because of the faster $T_2$ relaxation of brine in the W3 sample. Before the $N_2$ stream was turned on to initiate evaporation, an initial $^{23}$Na scan was taken. After this, a $^{23}$Na scan was taken every 1 hour three times, then every 2 hours twice, after which the interval was 6 hours until the end of the experiment.

Table 2. Key parameters of the imaging protocols used for $^1$H and $^{23}$Na MRI during evaporation of F36 and W3 samples. Two different protocols were needed due to the faster $T_2$ relaxation of brine in the W3 sample. MSSE: multi-slice single echo imaging using Bruker´s MSME protocol with a single Hahn echo; 3D RARE: three-dimensional rapid acquisition with relaxation enhancement.

| Scan Type | pulse seq. | $t_e$ (ms) | $t_r$ (ms) | slice | slice width (mm) | no. of avg. | time (min) | matrix size | resolution (mm) |
|---|---|---|---|---|---|---|---|---|---|
| $^1$H | MSSE | 4 | 3000 | 40 | 1 | 4 | 38 | 256×192×1 | 0.25×0.25×0.25 |
| F36 $^{23}$Na | 3D RARE | 10.2 | 300 | 1 | 40 | 8 | 26 | 64×64×40 | 1×1×1 |
| W3 $^{23}$Na | MSSE | 5.1 | 300 | 1 | 44 | 8 | 64 | 64×40×40 | 1×1.1×1.1 |

To verify that the samples remained saturated throughout the evaporation experiment, additional $^1$H-MRI scans were performed at the beginning, midpoint and end of the experiment. Water content was calculated by normalization on the intensity of a marker tube (see Figure 1a), which contained a mixture of 70% $D_2O$ and 30% $H_2O$. As $D_2O$ remains invisible when the probe head is tuned to the Larmor frequency of $^1$H, this mixture is representative for a sample with a volumetric water content of 0.30 cm$^3$ cm$^{-3}$. The volumetric water content ($\theta_{vol}$) of the packing was calculated using:

$$\theta_{vol} = \frac{S_{sample}}{S_{marker}} \cdot \varphi_{marker} \qquad (5)$$

where $S_{sample}$ is the mean signal intensity of $^{1}H$ for a given voxel area, $S_{marker}$ is the mean signal intensity in the marker tube, and $\varphi_{marker}$ is the volumetric fraction of $H_2O$ in the marker tube.

The Na concentration obtained with $^{23}Na$ MRI was calculated with the aid of a series of calibration samples scanned with the MRI protocols provided in Table 2. For this, F36 and W3 packings were prepared with known NaCl concentrations of 0, 1, 2, 3 and 5 mol $L^{-1}$ in the same sample holders using the packing method described above. The packings were covered to avoid evaporation. The Na volume concentration ($c_{Na,vol}$) of each calibration sample was calculated using (Perelman et al., 2020):

$$c_{(Na.vol)} = \frac{n_{Na}}{V_{bulk}} = c_{(Na.sol)} \cdot \theta_{vol} \qquad (6)$$

where $n_{Na}$ is the number of moles of Na, $V_{bulk}$ is the bulk volume, $c_{Na.sol}$ is the solution concentration and $\theta_{vol}$ is the volumetric water content. The average voxel intensity of the calibration samples was related to the Na volume concentration using a linear and a polynomial function for the F36 and W3 samples, respectively. The fitted calibration functions were then used to calculate Na concentration maps from the MRI signal intensity maps obtained during evaporation.

MRI data processing involved two steps. First, raw data were processed by inverse Fourier transformation within Paravision. The resulting images were further processed using ImageJ (version 1.53t). For enhanced visualization, all scans were equally scaled using the brightness/contrast function and then set to a standardized brightness range within the used color table. To reduce noise, a 1.5 pixel median filter was applied.

It is worth noting that MRI for $^{23}Na$ may be affected by a lower signal to noise (SNR) ratio due to the lower volume concentration and the much smaller gyromagnetic ratio compared to $^{1}H$. This is compensated by a higher number of accumulation, facilitated by short $T_1$ relaxation time. In addition, the fast $T_2$ relaxation of $^{23}Na$ may lead to $T_2$ weighted images if the echo time can not be set sufficiently short (Perelman et al., 2020).

### 3.3 Numerical simulation

The redistribution of solutes during evaporation from saturated porous medium was also investigated using numerical simulations. The mathematical model describing these processes is detailed in Appendix A. This model is implemented using the open-source simulator and research code DuMux (short for *DUNE for Multi-Phase, Component, Scale, Physics, ... flow and transport in porous media*) (Koch et al., 2021). A two-dimensional bounded domain Ω (in vertical direction between $z$ = 0 and $z$ = h, and in horizontal direction between $x$ = 0 and $x$ = d) was used. A cell-centered control volume method was used for spatial discretization and an implicit Euler scheme was adopted for the temporal discretization. In the vertical direction, the domain was discretized using a grid cell height Δz= 0.001 m with a grid refinement towards the top of the domain. In the horizontal direction, the grid length was Δx= 0.001 m. The temporal discretization was Δt = 600 s.

On the left and on the right boundary, the PMMA wall of the sample holder was mimicked by assuming that there is energy transfer but no mass transfer. At the top boundary, we assumed a constant evaporative flux $E$, while for salt a no flux condition was applied. The applied evaporation rate was the mean evaporation rate from the experiments ($E$ = 6 mm d$^{-1}$ for F36; $E$ = 4.5 mm d$^{-1}$ for W3). To allow heat loss due to evaporation, a heat flux across the interface between the porous medium and the atmosphere was considered. At the bottom boundary, no-flow conditions were assumed for mass transport. For energy transport, the same conditions were used as for the left and right boundary. To ensure the supply of water from below as in the experiment, a Nitsche-type boundary condition was applied at the water inlet on the bottom boundary. A description of this boundary condition is provided in Appendix A.

The initial conditions of the model simulations were a temperature of 293.15 K, a salt mole fraction of 0.0177 mol mol$^{-1}$ (concentration of 1 mol L$^{-1}$) and a gas pressure of 10$^5$ Pa. To trigger the density-driven instabilities, small imperfections were introduced by varying the salt mole fraction over the domain. For each cell, a random variation in salt mole fraction $x_{l,p}^s$ was added based on a normal distribution $\mathcal{N}$ with a mean mole fraction $x_l^s = x_{l,0}^s$ and a standard deviation $\sigma$ of 10$^{-5}$ mol mol$^{-1}$:

$$x_{l,p}^s \sim \mathcal{N}(x_l^s\,, \sigma) \quad (7)$$

The hydraulic parameters used for the simulations are listed in Table 1. For W3, the saturated water content was set to $\theta_{sat} = 0.36$ cm$^3$ cm$^{-3}$ to match the measured water content of the W3 sample used for the MRI investigations.

## 4. Results and Discussion

### 4.1 Evaporation dynamics

The cumulative water loss over time for the F36 and W3 samples is shown in Figure 2a. Figure 2b presents the associated evaporation rates for the two samples and the mean volumetric water content in the top 2 mm of each sample at different times obtained with $^1$H MRI. It was observed that $\theta_{vol}$ values were constant at 0.40 cm$^3$ cm$^{-3}$ for F36 and 0.36 cm$^3$ cm$^{-3}$ for W3. Using the constant nitrogen flow rate of 60 dm$^3$ h$^{-1}$, the average evaporation rate was approximately 6 mm d$^{-1}$ for F36 and this rate decreased only slightly until the end of the experiment at 121 hours (Figure 2b). In contrast, the W3 sample showed a sharp decrease in evaporation rate from nearly 9 mm·d$^{-1}$ for the first 2 to 3 hours to 4 mm·d$^{-1}$ at the end of the experiment. This initially higher evaporation rate is attributed to the presence of a thin water film that developed due to the movement of the sample to the MRI and associated minor settling. No visible salt crust formation was observed at the end of the experiments for both samples.

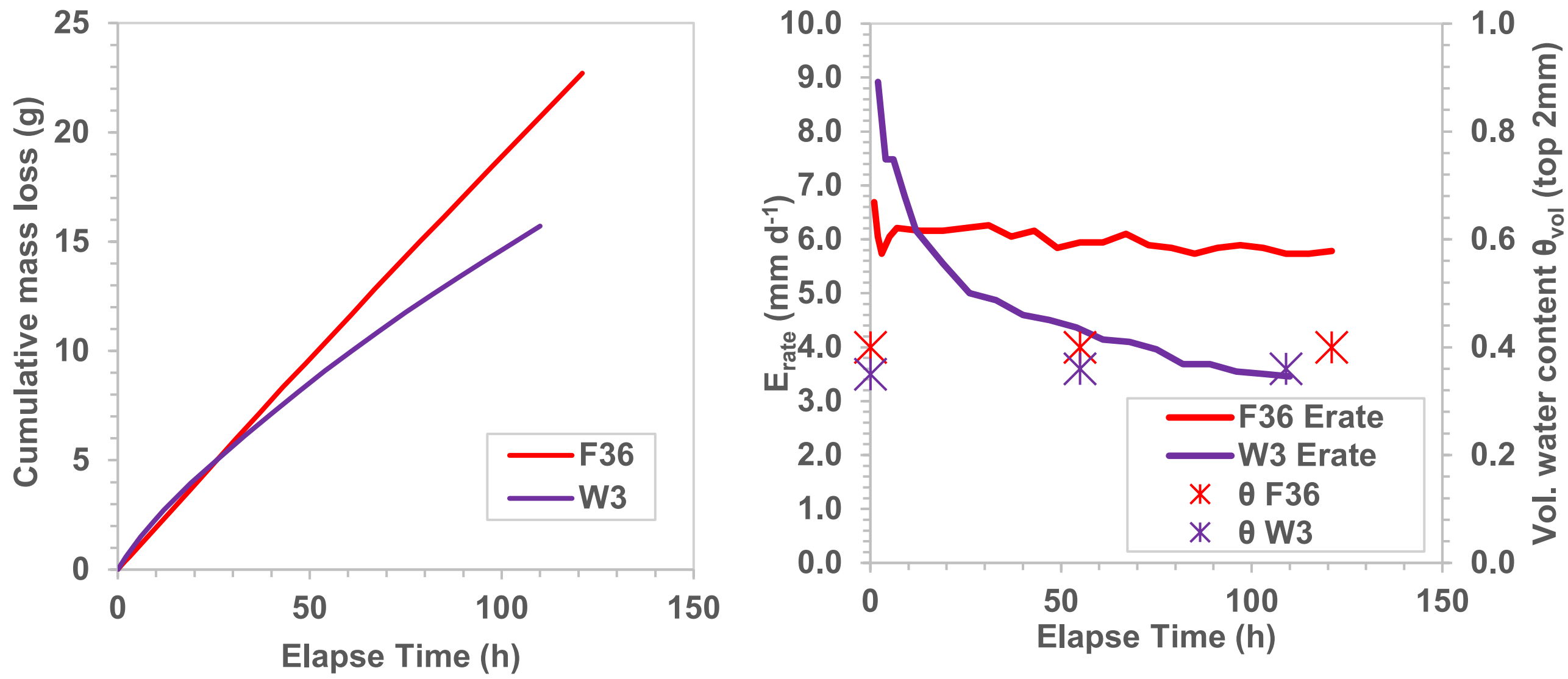

(a) (b)

Figure 2. (a) Cumulative mass loss of water during the evaporation experiments for the F36 and W3 samples. Both sands were initially saturated with 1 mol $L^{-1}$ NaCl solution and during evaporation 1 mol $L^{-1}$ NaCl solution was supplied from an outer reservoir. (b) Evaporation rate ($E_{rate}$) according to Equation 3 for the F36 and W3 samples (left ordinate) and volumetric water content $\theta_{vol}$ in the top 2 mm at the beginning, middle and end of the experiment (right ordinate).

The surface water content in the top 2 mm remained constant throughout the experiment and no dry surface was observed (Figure 2b). Therefore, it can safely be assumed that both samples experience stage 1 evaporation conditions with the evaporation front at the sample surface (Lehmann et al., 2008). As no salt precipitation was observed, all changes in evaporation rate need to be attributed to differences in the gradient of vapor pressure that occur in stage SS1 evaporation of saline water. The evaporation rate developed differently for both samples, which suggests differences in solute enrichment near the surface. In particular, the stronger decrease in evaporation rate for W3 is indicative of high solute concentrations near the top of the sample, while the higher evaporation rate for F36 suggests lower solute concentrations.

### 4.2 Calibration of $^{23}$Na MRI

To calculate Na concentration from measured MRI intensity, calibration relationships were established. Figure 3 shows the average voxel intensity of the calibration samples as a function of the Na volume concentration. During preparatory scans, bulk relaxation times of Na in the F36 sample were in good agreement with the relaxation times of Na in bulk solution approximating 50 ms at 4.7 T (Rijniers et al., 2004). A $T_2$ relaxation time of 35 ms was observed for F36 under fully saturated conditions, indicating minor contribution of surface relaxation due to the large pore sizes of F36. This bulk-like relaxation of Na in medium sand results in a linear relationship between $Na^+$(aq) concentration and MRI intensity, which was also observed by Perelman et al. (2020) when they performed calibrations for a similar sand at lower Na concentrations. The results presented here confirm that such bulk-like behavior is still applicable for higher concentrations, which was important to verify because of possible changes in hydration states of $Na^+$(aq) with increasing concentration (Rijniers et al., 2004). For W3, a non-linear calibration relationship was observed (Figure 3). For a fully saturated W3 sample, a $T_2$ value of 13.5 ms was observed for $Na^+$(aq), which indicates a considerable contribution of surface relaxation. This is attributed to the higher surface to volume ratio, which is estimated to be an order of magnitude higher for W3 based on the expected mean pore size estimated from water retention data. Surface relaxation of $Na^+$(aq) is attributed to a coupling of the quadrupole moment of $Na^+$(aq) with the electric double layer at the pore surface (Rijniers et al., 2004). Thus, we hypothesize that the unexpected increase of S(r) with concentration is due to stronger shielding of the surface charge at increased ionic strength.

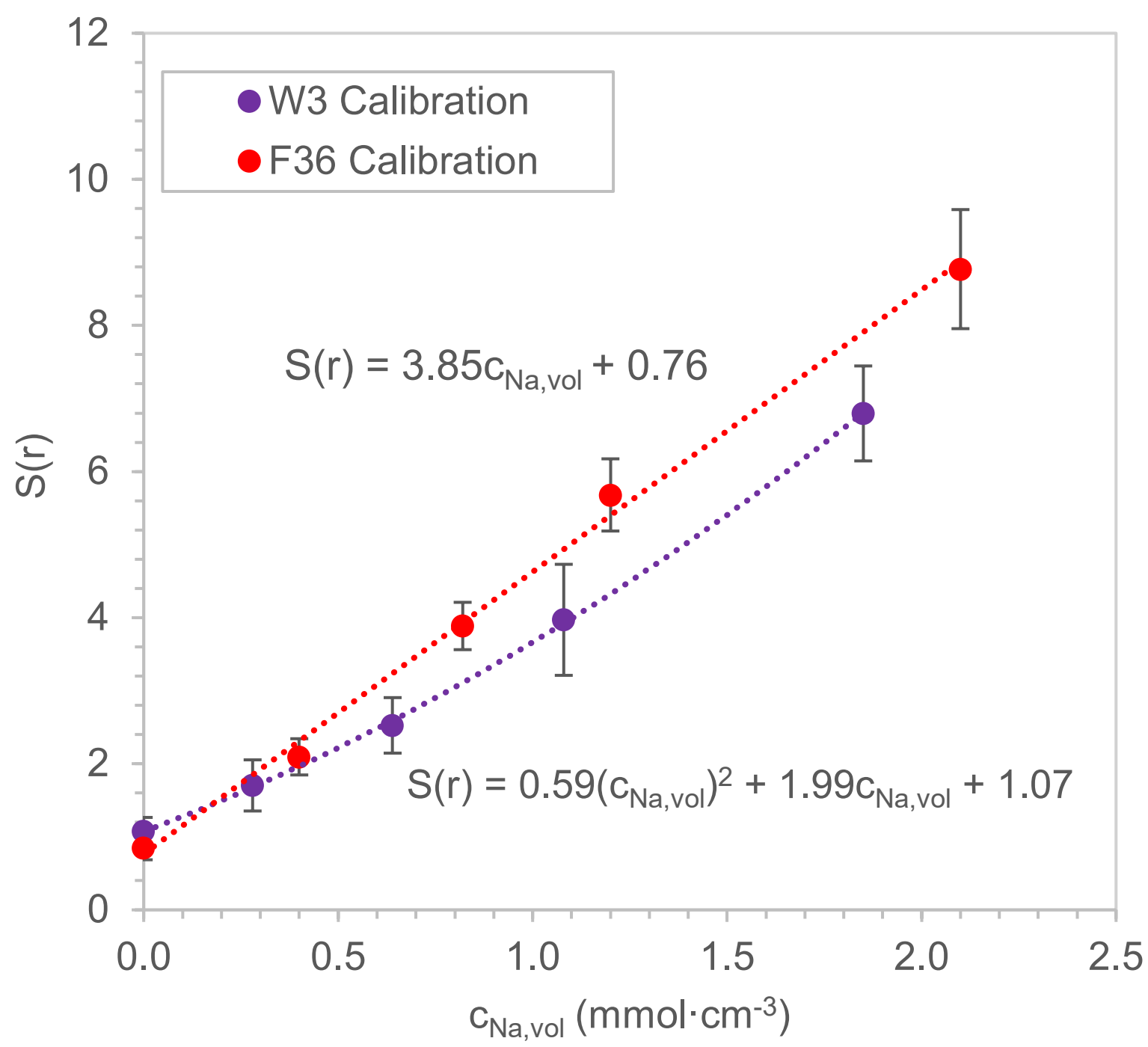

Figure 3. Calibration results for F36 and W3 where average voxel signal intensity ($S_r$) is related to the volume concentration of sodium ($c_{Na,vol}$) obtained with Equation 6. The dotted lines are a linear and a 2nd order polynomial fit with the parameters shown in the plot. The error bars represent 1 standard deviation of voxel intensity after using a 1.5 pixel median filter.

Figure 3 shows that the relatively low SNR for $^{23}$Na-MRI leads to substantial variation in voxel signal intensity even for the homogeneously packed calibration samples. To quantify the resulting uncertainty in Na concentration, the obtained calibration relationships were used to estimate the Na concentrations of the calibration samples. Assuming that all observed variation is due to the low SNR, the standard deviation of the concentration in the total voxel area of each sample (after filtering) can be used as an estimate of uncertainty in estimated concentration. For the F36 calibration series (0, 1, 2, 3 and 5 mol $L^{-1}$), the obtained standard deviations were 0.11, 0.16, 0.20, 0.32 and 0.51 mol $L^{-1}$. This indicates that the relative error in concentration is about 10%, which slightly increases with lower concentrations. For W3, the obtained standard deviations were 0.31, 0.54, 0.44, 0.64 and 0.40 mol $L^{-1}$. This indicates that the relative errors are higher for W3, except for the highest concentration. Here, it is important to note that this approach to error characterization is a worst case scenario, since all actual heterogeneity in packing density is considered within our error estimate. This likely at least partly explains the higher relative error for W3, which is much more difficult to handle during sample preparation.

### 4.3 Na enrichment patterns during evaporation

Figure 4 shows the spatial and temporal development of the Na concentration for the W3 sample, which is the low permeability medium in this study. A gradual enrichment of Na near the surface was observed with the sharpest concentration increase in the top 5 mm. Towards the end of the experiment at 103 h, the Na concentration reached a value close to the solubility limit. To build confidence in the imaging results and the selected calibration approach, the total amount of Na introduced into the column and the total amount of Na recovered from MRI was calculated as a function of time (Figure 5a). Good agreement was observed, and the difference in mass was consistently below 10% throughout the experiment despite the challenges with $^{23}$Na-MRI discussed previously. Errors were highest towards the end of the experiment, perhaps due to the occurrence of very high concentrations not considered during calibration. Figure 5b shows the temporal development of the mean vertical concentration profile of Na obtained from the imaging results in Figure 4. The concentration profiles below 36 mm have been linearly extrapolated from data points above this depth due to an observed packing heterogeneity. The concentration for the top 1 mm layer at 103 h is higher than expected (6.7 Mol $L^{-1}$), as it exceeds the solubility limit of 6.14 mol $L^{-1}$. However, supersaturation of the solution due to delayed nucleation cannot be excluded.

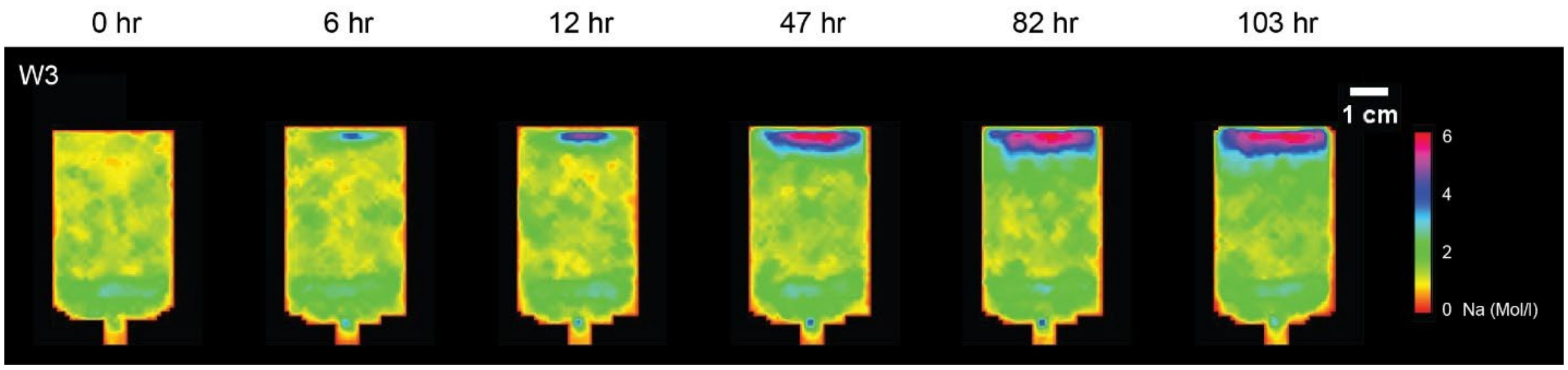

Figure 4. Time series of MRI images of Na concentration in a vertical cross section through the W3 sample for a central slice of 1.1 mm thickness during the evaporation experiment. The in-plane resolution was 1 mm x 1 mm. The time from start is indicated at the top of each image. Additional MRI parameters are summarized in Table 2.

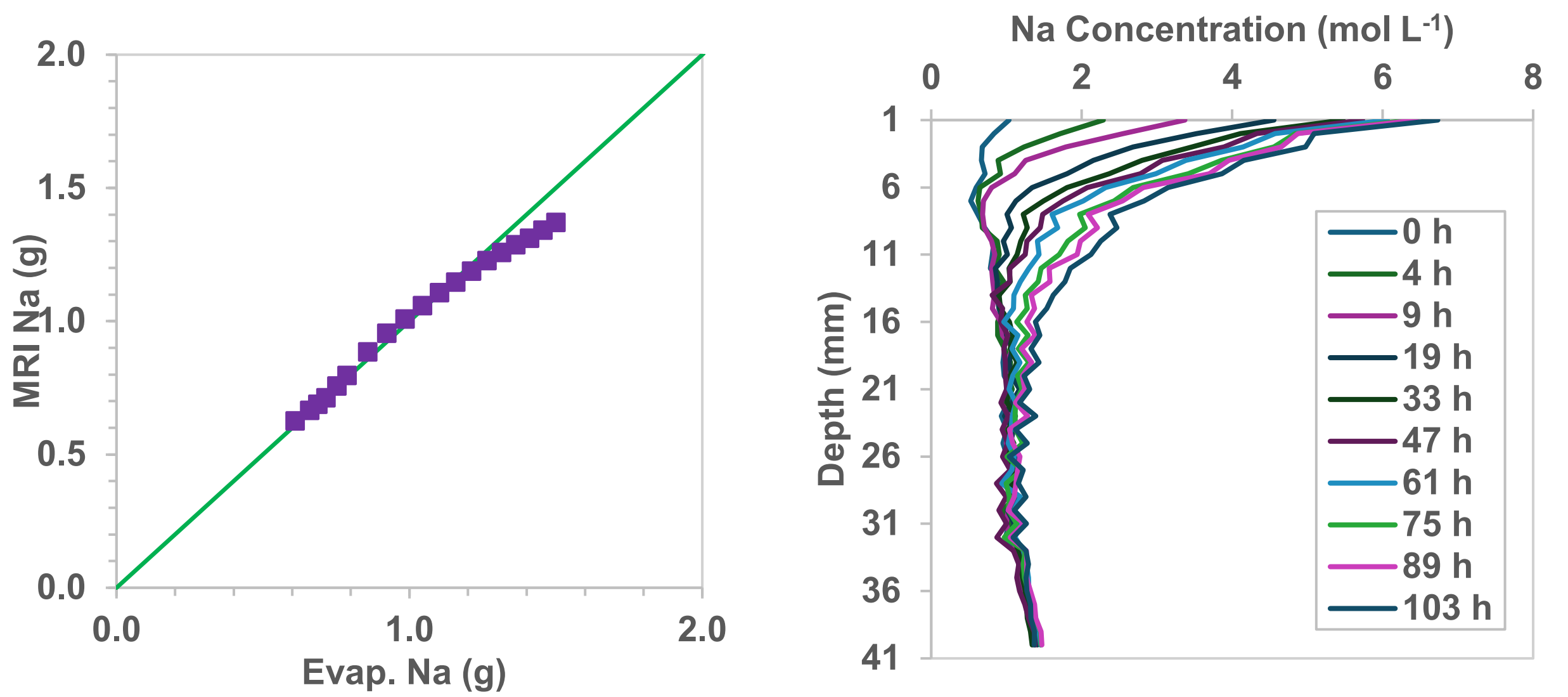

(a) (b)

Figure 5. (a) The calculated amount of Na from $^{23}$Na-MRI plotted against the total amount of Na in the column calculated from the mass loss data registered by the mass balance. The green line represents the 1:1 relation. (b) Depth profiles of Na concentration for the W3 sample during evaporation as a function of time.

Figure 6 shows the spatial and temporal development of Na concentration for the F36 sample obtained using $^{23}$Na MRI. Compared to the W3 sample, a distinctly different enrichment of Na was observed in the F36 sample, which has an intrinsic permeability that was two orders of magnitude higher than the W3 sample. Initially, the Na concentration increased at the surface as in the case of the W3 sample, but then a plume of higher concentration was seen moving downwards within the first 2 hours of evaporation. A continuous and intensifying downward flow of Na was seen to last throughout the duration of the experiment. The observed spatial variation in concentration far exceeded the estimated relative error of 10% in concentration, supporting the usefulness of $^{23}$Na-MRI for imaging unstable flows. As for the W3 sample, the mass balance of Na was evaluated (Figure 7a). A strong agreement between the amount determined with $^{23}$Na MRI and the amount of Na based on the evaporative mass loss was again observed, which inspires further confidence in the imaging results. The mean vertical Na concentration profiles for the F36 sample (Figure 7b) are also substantially different from the W3 sample. The observed downward movement of Na leads to increased concentration deeper in the sample, and an accumulation of Na in the bottom of the sample. The general behavior of density-driven downward flow against the evaporative flux described here is consistent with the theoretical considerations of Bringedal et al. (2022).

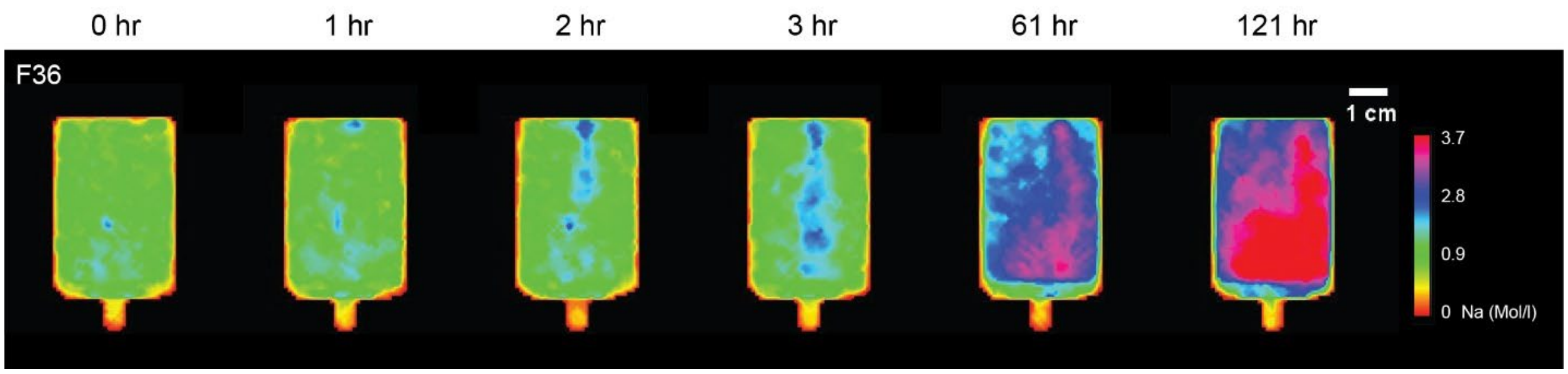

Figure 6. Time series of MRI images of Na concentration in a vertical cross section through the F36 sample for a central slice of 1 mm thickness during the evaporation experiment. The in-plane resolution was 1 mm x 1 mm. The time from start is indicated at the top of each image. The blue dot near the center of the column is an imaging artifact and may be ignored. Additional MRI parameters are summarized in Table 2.

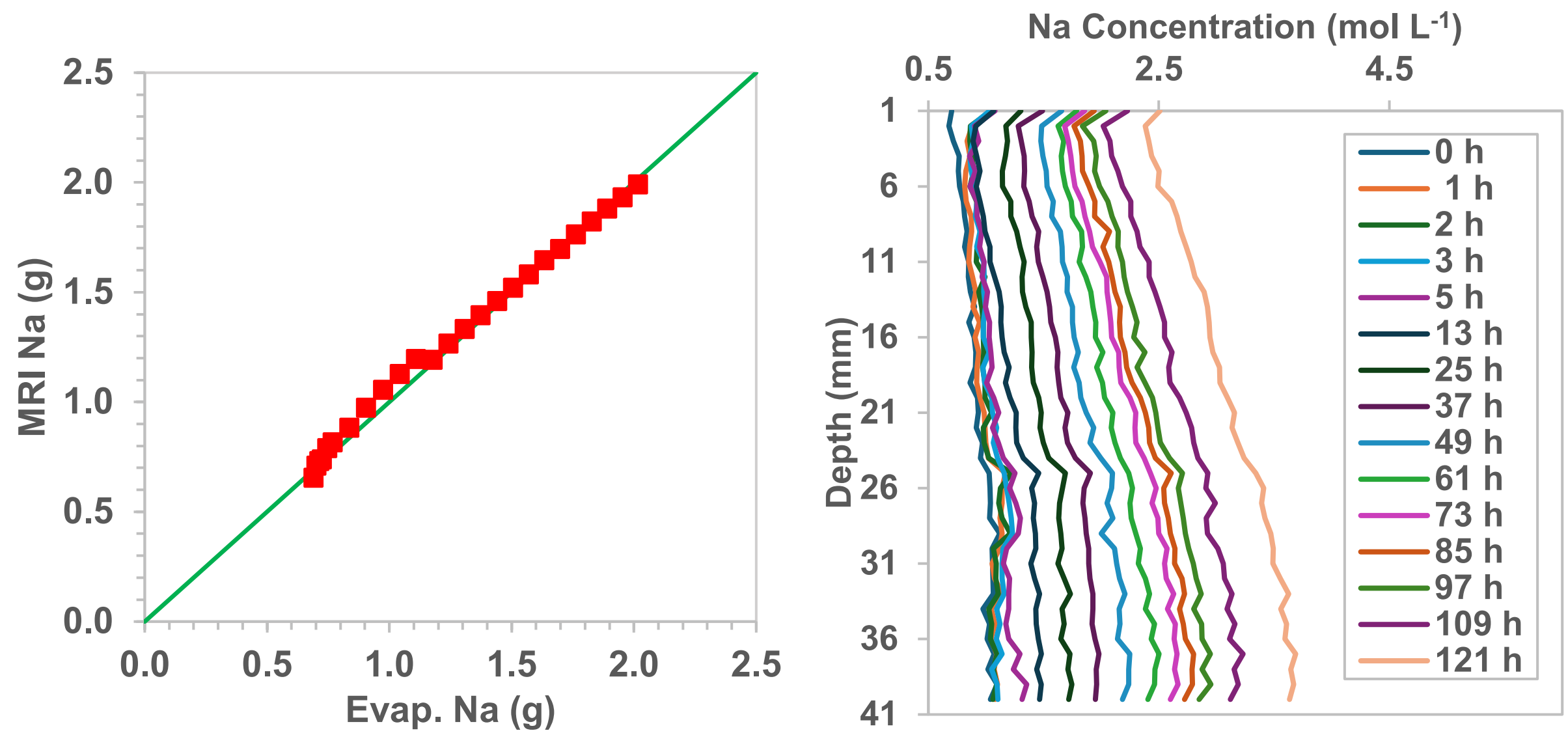

(a) (b)

Figure 7 (a) The calculated amount of Na from Na-MRI plotted against the total amount of Na in the column calculated from the mass loss data registered by the mass balance. The green line represents the 1:1 relation. (b) Depth profiles of Na concentration for the W3 sample during evaporation as a function of time.

The observed differences in the development of Na concentration are consistent with the difference in evaporation rate presented in Figure 2b. For the W3 sample, the solution concentration at the surface sharply increased (Figure 5b), which results in a lower saturation vapor pressure and a stronger decrease of the evaporation rate as typical for stage SS1 evaporation (Ho et al., 2006). Nachshon et al. (2011) also observed decreasing evaporation rates for homogeneous porous media undergoing saline water evaporation. For the F36 sample, a gradual increase of the Na concentration at the surface was observed (Figure 7b) due to the downward redistribution of Na. Even though concentration values at the surface of the F36 column increased to 2.4 mol $L^{-1}$, they were well below the solubility limit of 6.14 mol $L^{-1}$. Since the Na concentration profiles show only a slight relative increase at the surface of F36, the effect on the evaporation rate is lower than for the W3 sample.

MRI imaging showed an increase in Na concentration throughout the depth of the F36 sample due to the redistribution of NaCl (Figure 6). The analytical model of Bringedal et al. (2022) predicts that the highest concentrations occurred at the top of the porous media domain independent of the intrinsic permeability and the strength of the downward redistribution. The first 4 profiles (time =

0, 1, 2, 3 h) are consistent with these findings. However, once the fingering flow reaches the bottom of the sample holder, this leads to an accumulation of Na at the bottom of the sample. Therefore, the late-time differences in the development of the Na concentration profile between the experiments presented here and the analytical model of Bringedal et al. (2022) are attributed to the limited size of the sample holder that was not considered in the derivation of the analytical model.

Although $^{23}$Na-MRI provided useful insights into the distribution of Na, some issues with the imaging results beyond the signal-to-noise ratio that was already discussed can also be observed. For example, the initial concentration of 1 mol $L^{-1}$ was underestimated by $^{23}$Na-MRI in the top few millimeters of the F36 sample (Figure 7b). This is attributed to signal decay near the top of the field-of-view of the MRI rf coil. In the case of W3, the empirical polynomial calibration function resulting in a relatively good mass balance with a maximum deviation of only 8.7%. Nevertheless, concentrations higher than the solubility limit were observed near the surface, which may be due to the increased uncertainty in the non-linear calibration relationship for high concentrations. In addition, although no crust was seen upon sample removal, small quantities of crystalized salt may still be present that cannot be seen by $^{23}$Na-MRI. This may explain the minor underestimation of Na mass by the MRI towards the end of the experiment (Figure 5a).

**4.4 Stability analysis for experimental results**

It is of interest to compare the experimental results of this study with previously presented results on the stability of evaporating systems with near-surface solute enrichment. Wooding et al. (1997a) defined the boundary layer Rayleigh number ($R_\delta$) as:

$$R_\delta = \frac{\gamma g k}{v_r E_{rate}} \quad (8)$$

with

$$\gamma = \frac{\rho_i - \rho_{sol}}{\rho_{sol}} \quad (9)$$

where $v_r$ is the kinematic viscosity (assumed to be $1.05 \times 10^{-6}$ $m^2s^{-1}$ and $1.4\times 10^{-6}$ $m^2s^{-1}$ for 2 and 6 mol $L^{-1}$ NaCl solutions, respectively), $\rho_i$ is the density of the solution at the time that instability occurs, and $\rho_{sol}$ is the background density (of 1 mol $L^{-1}$ NaCl solution). Based on stability considerations and experimental results, they argued that instabilities are likely to occur when $R_\delta$ is larger than ~10. For the F36 sand, instability was observed when the surface concentration reached 2 mol $L^{-1}$ ($\gamma = 0.0318$). Using an evaporation rate of 6 mm $d^{-1}$ and the intrinsic permeability given in Table 1, this results in a $R_\delta$ of 33.5. For W3, no instability was observed, so the highest possible density at the solubility limit was assumed at the surface ($\gamma = 0.155$). Using the mean evaporation of 5.13 mm $d^{-1}$ and the known permeability, this results in a $R_\delta$ of 0.59. Thus, the observed behavior for the two evaporating systems investigated here are in general agreement with the stability analysis of Wooding et al. (1997a).

Bringedal et al. (2022) presented a more advanced linear stability analysis extending the work of Wooding et al. (1997a). Using their analytical model, we predicted the onset time of the instability formation and compared it with the experimental results. For the F36 sample, the onset time for the formation of instabilities is 0.61 h using an average evaporation rate of 6 mm $d^{-1}$. This is in good agreement with the time at which the first finger was visible in the MRI results (1 h), which we consider to be a reasonable proxy for the start of instability formation. For the W3 sample, a much longer onset time of 555 h was predicted using an average evaporation rate of 5.13 mm $d^{-1}$. This is consistent with the lack of instabilities in the experimental results.

**4.5 Comparison with numerical simulations**

In a final step, the experimentally observed solute distributions for the two samples with contrasting permeability were also compared to 2D numerical simulations – albeit in a qualitative manner only. Figure 8 shows the simulated concentration distribution at four different times for both materials, and Figure 9 shows the simulated vertical average concentration profiles. The simulations were initiated with a randomly perturbed concentration distribution with a mean of 1 mol $L^{-1}$ (Figure 8a,b). These perturbations represent small heterogeneities in the samples and are required to trigger flow instabilities. The temporal development of the simulated concentration profiles is generally consistent with the MRI imaging results. For the W3 sample, an increase in solute concentration is simulated with the highest concentrations at the top of the domain (Figure 9a). These simulated concentration profiles are consistent with the modelling results of Elrick et al. (1994) for an evaporating sample with a constant water content and also match well with the experimental results (Figure 5b). It should be noted that the simulation was terminated after 72 hours because the solubility limit was reached in the near-surface region and the resulting precipitation clogged the pore space and reduced the permeability to such an extent that the prescribed evaporation flux could not be maintained anymore.

For the F36 sample, the simulated solute distribution also shows initial enrichment at the top of the domain, but the system becomes gravitationally unstable with progressing time (Figure 8b). This results in the development of two fingers at the top of the domain after 11 hours. These fingers merge into one finger at the middle of the domain after 20 hours. A single finger in the middle of the domain was also observed in the experiment. The simulated finger moves further down along the vertical axis until reaching the bottom of the domain, after which it moves to the left side of the domain. Subsequently, salt is transported downwards on the left side and upwards on the other side of the domain. The movement of the finger towards the wall of the column was also observed in the experiments. The downward redistribution of solute results in a substantial accumulation of Na in the lower part of the domain (Figure 9b), which is consistent with the experimental results (Figure 5b). In contrast to the experimental results, the modelled concentrations show a maximum concentration at the top of the column, especially in the left corner of the domain (Figure 8b).

Although stochastic perturbations are needed to initiate density-driven instabilities, the formation of density-driven downward transport in the middle of the domain may not only be a coincidence. Jannesarahmadi et al. (2024) showed that a temperature gradient is likely to exist between the rim and center of an evaporating cylindrical sample with temperatures near the rim being 1° - 2° C warmer. By assuming the same temperature pattern on our sample surface, saline water near the rim would be slightly less dense than saline water in the center. Therefore, it seems likely that density driven downward flow is initiated in the center of the sample in our experimental set-up.

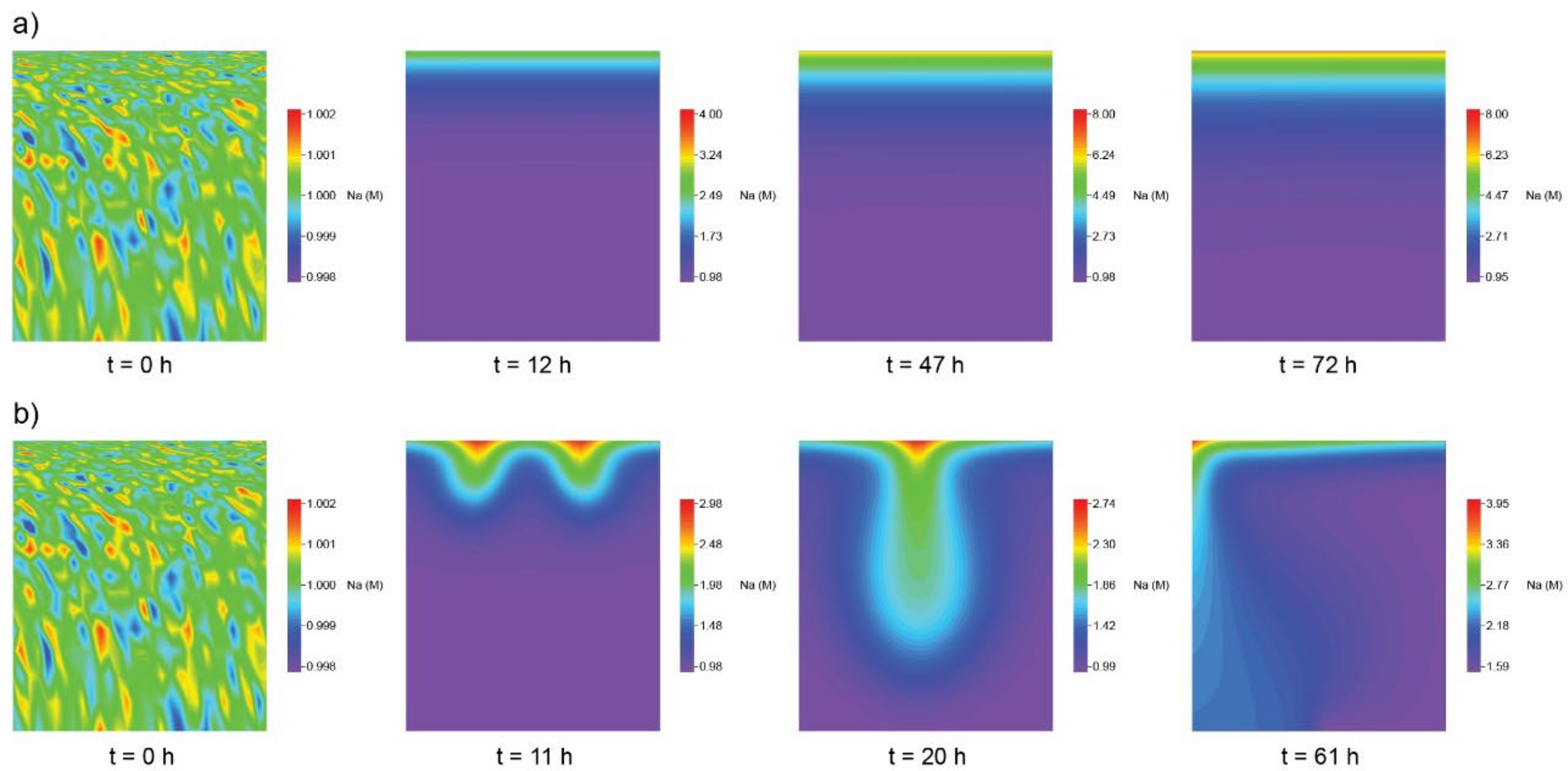

Figure 8. Results of a 2D numerical simulation of salt concentration development. Distribution of the Na concentration across the entire domain (3.1 cm × 4 cm) at different points in time of the simulations for the (a) W3 sample and the (b) F36 sample.

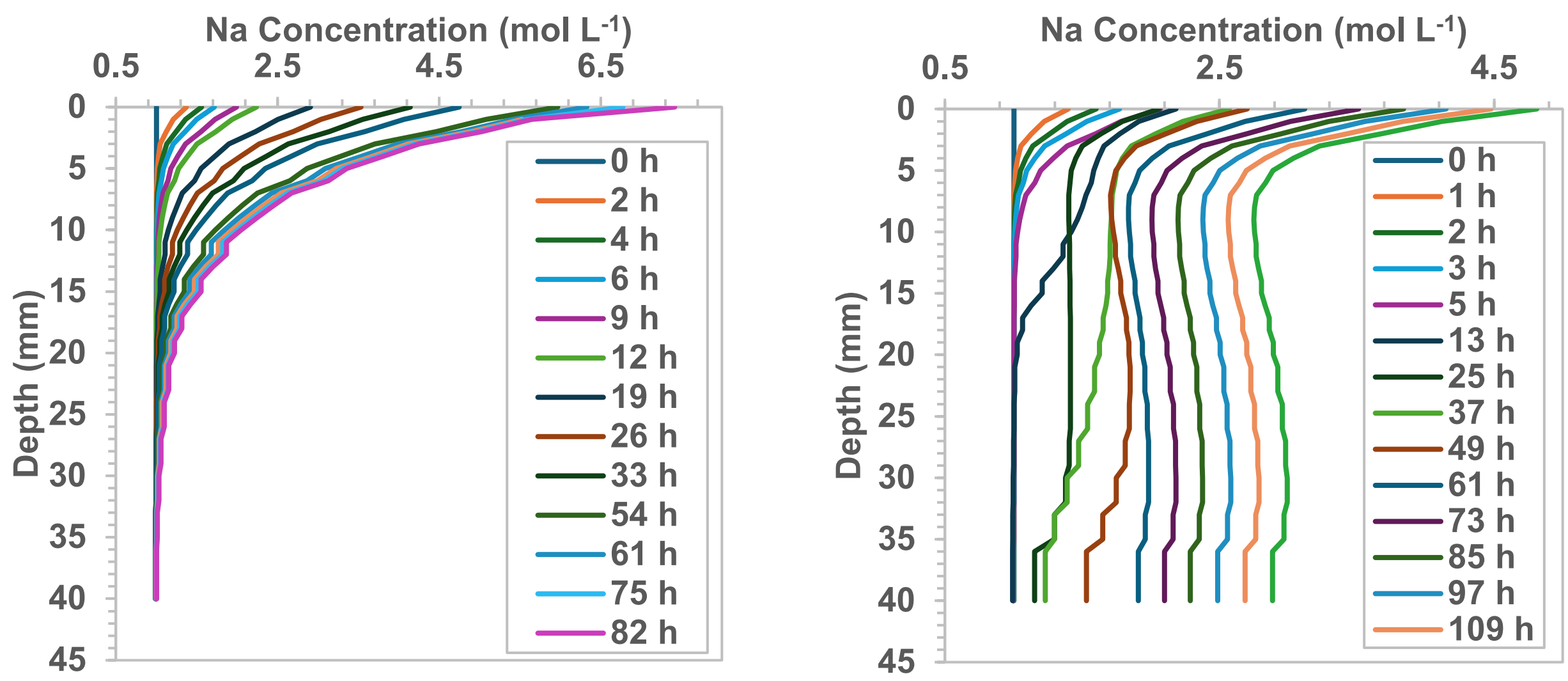

Figure 9. Average vertical Na concentration over the domain height at different times of the simulations for (a) W3 and (b) F36.

The model simulations can also be used to determine the characteristic wavelength of instability to verify that the selected sample width had no impact on the formation of instabilities. Slim (2014) described regimes of instability growth from initial diffusive flow to a linear growth regime where faint perturbations first become apparent. The time it takes to arrive at this stage is defined as linear-onset time and can be used to determine the wavelength (λ) of the first instability. The linear-onset time obtained from the numerical modelling results for F36 predicted a wavelength of 0.01 m, which is smaller than our column diameter (0.031 m). The shift from linear growth regime to the flux growth regime is marked by mass transport and can be considered as the initiation of finger formation. The estimated wavelength value at this transition was 0.0155 m. This suggest that instability formation is independent of sample width. Due to limited length of the samples investigated here, the late time results are affected by the lower boundary.

In summary, the numerical simulations were able to match several key features of the experimental data, including the observed differences in density-driven downward flow leading to largely different vertical concentration profiles for the two investigated porous media. Due to the inevitable simplifying assumptions, some differences were also observed. For example, the numerical model does not exactly reproduce the amount of salt in the system at all times due to the simplified upper boundary conditions with a prescribed and constant evaporation flux. In addition, the simulated Na concentration reached the solubility limit earlier than observed for W3, which is attributed to the uncertainty in the hydraulic properties. For the F36 sample, the model simulates the highest concentrations at the top of the domain, but a considerable enrichment at the bottom is also simulated. This is only partly consistent with the experimental results, which showed a stronger enrichment near the bottom of the sample. This is attributed to the difficulty of representing the lower boundary in the model. An extension from a two-dimensional model to a three-dimensional model could possibly enhance the comparison, as the flux is larger in a three-dimensional system (Liyanage et al., 2024).

## 5. Conclusions and Outlook

In this study, we investigated solute redistribution (NaCl) for two evaporating porous media differing in intrinsic permeability under fully saturated conditions. Time-lapse imaging of the 3D distribution of Na concentration with $^{23}$Na-MRI demonstrated that the development of the solute concentrations differed substantially depending on the porous media properties. In particular, a shallow near surface enrichment zone was observed for the low permeability sample ($3.2\times10^{-14}$ $m^2$) with Na concentrations reaching the saturation limit. Due to the strong increase of Na concentration at the surface, a substantial decrease in evaporation rate was observed for this sample. In case of the sample with higher permeability ($7.8\times10^{-12}$ $m^2$), an initial enrichment at the surface was also observed within the first hour, but soon after a downwards moving plume developed that redistributed NaCl back into the column. This was attributed to density-driven flow made possible by the high permeability. For this sample, the Na concentration at the sample surface only moderately increased and the observed evaporation was therefore fairly constant. A comparison of the experimental results with numerical simulations using $DuMu^x$ showed good qualitative agreement where fingering in the high-permeability sample was also observed to distribute salt along the depth of the column and the low-permeability sample showed only surface enrichment of salt concentration for the investigated experimental conditions.

The presented results showed that density-driven flow associated with evaporation is capable of redistributing solutes away from the evaporating surface. This phenomenon was predicted from analytical and numerical modelling (Wooding et al., 1997; Bringedal et. al., 2022) and has now been experimentally confirmed for an evaporating porous media column. The findings extend our understanding of the theory of saline water evaporation and provide novel insights into solute redistribution during evaporation in the presence of a shallow water table. Beyond evaporation, the work also carries implications for the complex hydrology associated with the formation of salt flats, which are typically accompanied by saline groundwater at depth (e.g. Hamann et al., 2015).

The results open up new directions for research related to the implications of density-driven instabilities on the associated solute redistribution, such as the required time to crust formation for different media undergoing saline water evaporation. Here, a faster production of crust is anticipated for low-permeability media in case of evaporation from a fully saturated sample. It is also of interest to further investigate solute distribution patterns for heterogeneously packed evaporating systems, where intricate interactions between high and low permeability regions are anticipated due to downward redistribution and lateral diffusion. Finally, it is also of interest to extend this study to different salt types such as $Na_2SO_4$ and $MgSO_4$ given that the different density of these solutions will have an impact on the generation of instabilities and the resulting solute redistribution.

**Open Research**

Data & Software Availability Statement

The experimental data for solute redistribution below evaporating surfaces using $^{23}$Na-MRI are made available through Chaudhry & Huisman (2025) and can be found in the data repository of the University of Stuttgart (DaRUS; doi: https://doi.org/10.18419/DARUS-5094). The code for the numerical simulations is made available in Kiemle (2025) and can be found in the data repository of the University of Stuttgart (DaRUS; doi: https://doi.org/10.18419/DARUS-4744).

**Acknowledgements**

This work was funded by the German Research Foundation (DFG) through project C05 of the Collaborative Research Center 1313 (Project Number 327154368 – SFB 1313). The authors would further like to thank Carina Bringedal for the help with the linear stability analysis and Odilia Esser for the help with the determination of the hydraulic properties.

**Conflict of Interest Statement**

The authors have no conflict of interest to disclose.

**Appendix A: Description of the mathematical model**

In the following, we describe the mathematical model and the underlying physical assumptions to describe evaporation from a porous medium saturated with saline water. We account for two phases α, the liquid phase l (water, air, dissolved salt) and the gas phase (water, air), to allow potential local unsaturated zones. Mass conservation is solved for the components $\kappa \in \{w, a, s\}$ and the phases $\alpha \in \{l, g\}$:

$$\frac{\partial(\Sigma_\alpha \phi \varrho_{\alpha,(m)} x_\alpha^\kappa S_\alpha)}{\partial t} + \nabla \cdot F^\kappa = \Sigma_\alpha q_\alpha^\kappa \ , \ F^\kappa = \Sigma_\alpha\left(\varrho_{\alpha,(m)} x_\alpha^\kappa v_\alpha - D_{\alpha,pm}^\kappa \varrho_{\alpha,(m)} \nabla x_\alpha^\kappa\right) \quad \text{(A1)}$$

with the porosity $\phi$, the molar phase density $\varrho_{\alpha,(m)}$, the mole fraction of component in phase $x_\alpha^\kappa$, and the source term $q^\kappa$. The volumetric water content is here defined as the phase saturation $S_\alpha := \frac{\theta_\alpha}{\phi}$. The effective diffusion coefficient $D_{\alpha,pm}^\kappa$ is obtained using the approach of Millington and Quirk (1961). The fluid phase velocity $v_\alpha$ is described using the multi-phase Darcy's law:

$$v_\alpha = \frac{k_{r,\alpha}}{\mu_\alpha} k(\nabla p_\alpha - \varrho_\alpha g_z) , \quad \text{(A2)}$$

with the relative phase permeability $k_{r,\alpha}$, the phase viscosity $\mu_\alpha$, the intrinsic permeability $k$, the phase pressure $p_\alpha$ and the gravity in vertical direction $g_z$.

As we are solving a partially saturated system, we need a closure relationship to derive the capillary pressure $p_c$. Here we use the Brooks-Corey-relationship:

$$S_e = -\frac{p_c}{p_b} S_e^{-\lambda_{BC}} . \quad \text{(A3)}$$

We describe the capillary pressure – saturation relationship, by using the effective saturation $Se := \frac{S_w - S_{wr}}{1 - S_{wr}}$, and two empirical parameters: the Brooks-Corey parameter $\lambda_{BC}$ and the entry pressure $p_b$. The residual water saturation is expressed by the residual water content $S_{wr} := \frac{\theta_{res}}{\phi}$.

Once the solute concentration exceeds the solubility limit $x_{l,max}^s$, here $x_{l,max}^s = 0.0977$ mol mol$^{-1}$, salt will precipitate and thus alter the porous medium. While $q^\kappa = 0$ for $\kappa \in \{w, a\}$, the source term for $\kappa \in \{s\}$ is described as

$$q^s = \phi \varrho_{l,(m)} S_l \left(x_l^s - x_{l,max}^s\right) . \quad \text{(A4)}$$

As salt precipitates, it accumulates within the porous medium over time, which needs to be considered to ensure mass conservation:

$$\frac{\partial\left(\phi_S^s\, \varrho^s{}_{(m),S}\right)}{\partial t} \;=\; q^s \tag{A5}$$

To account for the alteration of the pore space due to salt precipitation $\phi_S^s$, the change in porosity is then given by:

$$\phi \;=\; \phi_0 - \phi_S^s. \tag{A6}$$

Here, $\phi_0$ is the initial porosity. The permeability is then modified by using a Kozeny-Carman-approach:

$$\frac{k}{k_0} \;=\; \left(\frac{\phi}{\phi_0}\right)^3 \left(\frac{1-\phi_0}{1-\phi}\right)^2, \tag{A7}$$

with $k_0$ as the initial permeability.

Heat transport is described by

$$(1-\phi_0)\frac{\partial \varrho_S c_S T}{\partial t} + \frac{\partial \phi_S^s \varrho_S^s\, c_S^s\, T}{\partial t} \;+\; \Sigma_\alpha \frac{\phi \varrho_\alpha u_\alpha S_\alpha}{\partial t} \;+\; \nabla\cdot\, F^T \;=\; 0,\;\; F^T = \left(\Sigma_\alpha \varrho_\alpha h_\alpha v_\alpha \;-\; \lambda_{\,pm}\nabla T\right) \tag{A8}$$

assuming local thermodynamic equilibrium. Here, $c_s$ is the specific heat capacity of the porous medium and $c_S^s$ is the specific heat capacity of the precipitated salt. Further, the energy conservation is described with the fluid-specific internal energy $u_\alpha$, the temperature $T$, the specific enthalpy $h_\alpha$ and the effective thermal conductivity $\lambda_{pm}$.

For the numerical simulations, *x* describes the horizontal axis and *z* the vertical axis where both start on the lower left corner of the domain. We use three different boundary types depending on the position. For the Type I boundary condition at the top of the sand column, we allow the system to evaporate water and prescribe that salt remains within the porous medium:

$$F^k = 0 \;for\; k\epsilon\{a,s\}, F^w = E\; \varrho_{l,(m)}\left(T_{a,}p_g\right), \tag{A9}$$

$$F^T = H_g^w\left(T_{a,}p_g\right)F^w M^w \,+\, \lambda_\alpha\left(T_{a,}p_g\right)\frac{T_{pm}-T_a}{\delta}, \tag{A10}$$

with the evaporation rate $E$ (here: in m $s^{-1}$), the ambient air temperature of $T_a$ = 293.15 K, the specific enthalpy of the component $H_{\alpha'}^k$, the molar mass of the component $M^k$, and the thermal conductivity of the component air $\lambda_a$. The boundary layer thickness $\delta$ has been calculated based on the evaporation rate of the respective sand (δ(W3) = 0.003 m, δ(F36) = 0.013 m). The Type II boundary condition is used for the PMMA sides of the sample holder. There, we allow energy transfer but no mass transport:

$$F^k = 0 \; for \; \kappa \in \{w, a, s\}, \qquad F^T = \lambda_{PMMA} \frac{T_{pm} - T_a}{d_{PMMA}}, \tag{A11}$$

with the thermal conductivity $\lambda_{PMMA}$= 0.184 W m$^{-1}$ K$^{-1}$, the temperature inside the porous medium at the border of the domain $T_{pm}$, and the wall thickness of the polymethylmethacrylated$d_{PMMA}$ = 0.005 m. The Type III boundary condition is located at the inlet of the brine solution (for $z$ = 0, and for 0.0045 m ≤ $x$ ≤ 0.0065 m):

$$F^w = -\nabla.\left(\frac{k_{r,a}}{\mu_\alpha} K(\nabla p - \varrho_l g)\right), F^s = F^w x^s_{l.0}, F^T = F^w H^w_l(T_{pm}, p_g) M^w \tag{A12}$$

This boundary condition is based on a fixed bottom pressure ($p_{l,z=0} = p_g - z(=h)\varrho_l g$) and sets the Neuman flux calculated from the pressure difference, the mobility, and permeability. Thus, the boundary condition is weakly enforced by combining the pressure-driven flux with the variable flux.